\documentclass[journal]{IEEEtran}
\usepackage{mathrsfs}
\usepackage{bbm}
\usepackage{amssymb}
\usepackage{bbding}
\usepackage{threeparttable}
\usepackage[justification=centering]{caption}
\usepackage[mathcal]{euscript}

\usepackage{psfrag,calc,url,bm}

\usepackage{cite}

\usepackage{graphicx}

\usepackage{algorithm}
\usepackage{algorithmic}

\usepackage{psfrag}

\usepackage{subfigure}
\usepackage{ulem} 
\usepackage{url}
\usepackage{cite}
\usepackage{stfloats}

\usepackage{amsmath}

\usepackage{float}

\usepackage{color}

\usepackage{boxedminipage}

\usepackage{amsthm}

\usepackage{multirow}

\usepackage{setspace}

\usepackage{soul}
\usepackage{epstopdf}
\usepackage{ulem}
\newtheorem{remark}{Remark}

\allowdisplaybreaks[3]

\begin{document}

\title{BERT4beam: Large AI Model Enabled Generalized Beamforming Optimization}
\author{Yuhang Li, Yang Lu,~\IEEEmembership{Senior Member,~IEEE}, Wei Chen,~\IEEEmembership{Senior Member,~IEEE}, Bo Ai,~\IEEEmembership{Fellow,~IEEE}, \\Zhiguo  Ding,~\IEEEmembership{Fellow,~IEEE} 
\thanks{This work was supported in part by National Natural Science Foundation of China (NSFC) under Grant 62571023, U2468201, and 62221001, and in part by Beijing Natural Science Foundation under Grant L242086. {(Corresponding author: Yang Lu.)}} 
\thanks{Yuhang Li and Yang Lu  are with the State Key Laboratory of Advanced Rail Autonomous Operation, and also with the School of Computer Science and Technology, Beijing Jiaotong University, Beijing 100044, China (e-mail: 24110137@bjtu.edu.cn; 
 yanglu@bjtu.edu.cn).}
\thanks{Wei Chen and Bo Ai are with the School of Electronics and Information Engineering, Beijing Jiaotong University, Beijing 100044, China (e-mail: weich@bjtu.edu.cn; boai@bjtu.edu.cn).}
\thanks{Zhiguo Ding is with the School of Electrical and Electronic Engineering (EEE), Nanyang Technological University, Singapore 639798 (Zhiguo.ding@ntu.edu.sg).}
}
\maketitle

\begin{abstract} 
Artificial intelligence (AI) is anticipated to serve  as a pivotal enabler for the forthcoming sixth-generation (6G) wireless communication systems. However, existing research on large AI models for wireless communications primarily focuses on fine-tuning pre-trained large language models (LLMs) for specific tasks, lacking generalizability across diverse scenarios. This paper investigates the design of large-scale AI models for beamforming optimization, aiming to adapt and generalize across various tasks defined by system utilities and scales. We propose a novel framework based on bidirectional encoder representations from transformers (BERT), termed BERT4beam. We formulate the beamforming optimization problem as a token-level sequence learning task, tokenize the channel state information, construct BERT models, and implement task-specific pre-training and fine-tuning strategies. Based on the framework, we propose two BERT-based approaches for single-task and multi-task beamforming optimization, respectively, both of which exhibit generalizability across varying user scales. Furthermore, the former can adapt to varying system utilities and antenna configurations by reconfiguring the input and output modules of the BERT model, while the latter, termed UBERT, can directly generalize to diverse tasks via a finer-grained tokenization strategy. Extensive numerical results demonstrate that the two proposed approaches can achieve near-optimal performance and outperform existing AI models across various beamforming optimization tasks, showcasing strong adaptability and generalizability.

\end{abstract}


\section{Introduction}

The upcoming sixth-generation (6G) mobile communication system is expected to achieve ubiquitous and high-quality coverage for  a vast number of wireless devices. The development of smart radio enabler technologies has enhanced the availability of wireless resources, but it has also been confronted with challenges posed by the growing system complexities and diverse service demands \cite{6G}. In the past few decades, optimization methods based on convex (CVX) optimization theory have been employed as the primary techniques for addressing wireless signal processing problems\cite{2024:yafeng}. However, these methods typically require a significant amount of iterative time and are only effective when appropriate mathematical models are available. Therefore, from a practical perspective, traditional methods may be unsuitable for wireless networks with real-time dynamic changes\cite{2023:yifei}. The deep integration of artificial intelligence (AI) with wireless communication is becoming increasingly essential for 6G, and the AI models can serve as intelligent solvers for designing and optimizing wireless networks\cite{2019:KB,2024:shenzhe,2024:hongyang,2024:ruichen2}. 

The majority of existing works on deep learning (DL)-enabled wireless communications follow the ``learning-to-optimize" paradigm \cite{2018:haoran}, in which customized neural networks are trained to approximate traditional algorithms. However, with the advent of 6G, both the resource dimensionalities and the system scales expand dramatically. Consequently, it becomes increasingly difficult for small-sized models to capture generalized patterns for complicated application scenarios. Considering the diversity of wireless communication objectives and system configurations, there is an imperative need for a unified AI model capable of handling generalized tasks rather than only one specific task. Fortunately, large language models (LLMs) can be an attractive choice. The transformer \cite{Attention_Illia} architecture marked a significant milestone in the field of  natural language processing (NLP) and  AI, propelling LLMs to unprecedented levels of advancement. The success of transformer-based models such as bidirectional encoder representations from transformers (BERT)\cite{2019:bert}, generative pre-trained transformer (GPT)\cite{2018:GPT}, and the large language model meta AI (LLaMA)\cite{2023:LLaMA} series of LLMs demonstrates their remarkable generalization capabilities. Furthermore, the semi-supervised training strategy of pre-training and fine-tuning has garnered widespread attention. Some recent research endeavors  have been  dedicated to applying  LLMs in wireless networks, such as semantic communication\cite{2025:Huiqiang,2024:feibo}, edge computing\cite{2024:yifei,2025:ruichen}, resource management\cite{2025:Javaid}, and LLM-integrated systems\cite{2024:shenzhe,2023:Tarkoma,2024:ruichen}.  Compared to traditional AI models, LLMs, by  learning abstract representations from large-scale and diverse data, can significantly enhance their generalization and transfer abilities to adapt to various task requirements. 

{Despite the remarkable success of LLMs in NLP, their effective deployment in wireless communications remains non-trivial. Existing approaches often directly fine-tune pre-trained LLMs for wireless tasks, which gives rise to three limitations. First, LLMs are pre-trained on natural language data, which differs substantially from wireless-specific data such as CSI, undermining their adaptability and performance. Second, LLMs are inherently designed for textual data, whereas most wireless signals are numerical in nature, rendering direct application difficult. Third, prior works primarily focus on single-task optimization, lacking generalizability across tasks with diverse objectives or varying system scales. These challenges motivate the design of LLM frameworks tailored explicitly for wireless communications, which enable the processing of complex-valued signals and support multi-task learning to unlock the full potential of LLMs in wireless optimization problems.} To this end, this paper explores a unified AI model that leverages core  LLM technologies to address classical beamforming optimization problems. Specifically, we propose a novel BERT-based\footnote{BERT is adopted because its bidirectional attention mechanism is analogous to the core design of graph attention networks (GAT), which is widely used in wireless communications.} framework for beamforming optimization, termed BERT4beam, to handle multiple tasks with different system objectives and scales. The BERT4beam framework is devoted to developing LLMs for wireless networks in terms of CSI tokenization, BERT model construction, and dedicated pre-training and fine-tuning strategies. The main technical contributions are summarized as follows:
\begin{itemize}
\item {For single-task beamforming optimization, we propose a BERT-based approach. Specifically, the CSI of each user is tokenized and fed into a BERT model that is designed to process CSI tokens. Additionally, a supervised pre-training and fine-tuning strategy is developed based on a single-task loss function. The pre-trained BERT model exhibits strong generalization across different user scales, while fine-tuning enables efficient adaptation to practical deployment scenarios with mismatched user counts or varying antenna configurations.}

\item {For multi-task beamforming optimization, we develop a new model termed  UBERT. First, we design an element-wise tokenization strategy that maps channel gains to fine-grained tokens based on antenna–user pairs, thus enabling  the reuse of pre-trained weights and avoiding reinitialization during fine-tuning. Second, we integrate  an antenna encoding module and a task embedding layer into UBERT to fully exploit antenna-specific characteristics and differentiate among diverse tasks. Finally, we propose a supervised pre-training and fine-tuning strategy based on a multi-task loss function, coupled with a uniform task sampling scheme to improve training stability.}

\item {Extensive numerical experiments are carried out to validate the proposed approaches. The results demonstrate that both single-task BERT and multi-task UBERT models attain performance on par with conventional optimization methods across multiple beamforming tasks, while exhibiting superior adaptability and robustness in dynamically varying wireless scenarios.}
\end{itemize}

The rest of this paper is organized as follows. {Section II reviews both CVX-based wireless design and emerging AI-enabled wireless communications.} Section III introduces the problem definition and the concept of BERT4beam. Section IV and Section V detail two BERT-based models.  Section VI presents numerical results and evaluation. Section VII concludes the paper.  

{{\it Notation}: The following mathematical notations and symbols are used throughout this paper. Bold lowercase letters (e.g., $\mathbf{a}$) denote column vectors, and bold uppercase letters (e.g., $\mathbf{A}$) denote matrices or higher-dimensional tensors. The set of real numbers is denoted by $\mathbb{R}$, and the set of $n \times m$ real matrices by $\mathbb{R}^{n \times m}$. Similarly, $\mathbb{C}^n$ and $\mathbb{C}^{n \times m}$ denote the sets of $n$-dimensional complex column vectors and $n \times m$ complex matrices, respectively. For a complex number $a$, $|a|$ denotes its modulus, and $\Re(a)$ and $\Im(a)$ denote its real and imaginary parts, respectively. For a vector $\mathbf{a}$, $\| \mathbf{a} \|$ denotes its Euclidean norm. {For a matrix $\mathbf{A}$, $\mathbf{A}^T$, $\mathbf{A}^H$, $\| \mathbf{A} \|$, and $\text{Trace}(\mathbf{A})$ represent its transpose}, conjugate transpose, Frobenius norm, and trace (i.e., the sum of its diagonal elements), respectively. The notation $\mathbf{a}[i]$ denotes the $i$-th entry of vector $\mathbf{a}$; $\mathbf{A}[i,j]$ denotes the element in the $i$-th row and $j$-th column of matrix $\mathbf{A}$; and $\mathbf{A}[i,:]$ denotes the entire $i$-th row of $\mathbf{A}$. The operator $\text{Concat}(\cdot)$ denotes the concatenation of its input(s). 

}

\section{Related Works}

\subsection{{CVX-Based Wireless Design}}
{
Over the past decades, beamforming design has been predominantly addressed using traditional optimization algorithms. For instance, a successive convex approximation (SCA)-based beamforming algorithm was proposed in \cite{paper:shen} to improve the spectral efficiency of multi-user multiple-input single-output (MU-MISO) systems. In \cite{paper:Ya-Feng}, the authors employed the block coordinate descent (BCD) and gradient projection (GP) methods to address the coordinated beamforming problem in multi-cell networks. The widely recognized weighted minimum mean square error (WMMSE) algorithm, introduced in \cite{paper:Christensen} and \cite{paper:Qingjiang}, leverages the equivalence between the signal-to-interference-plus-noise ratio and mean square error (MSE) \cite{paper:Qingjiang} and adopts the BCD method to derive optimal solutions. Additionally, fractional programming has been proposed for energy efficiency (EE)-oriented beamforming design \cite{paper:lu,paper:li}. Nevertheless, these iterative methods exhibit inherent high computational complexity, which poses substantial challenges to real-time processing in practical time-varying wireless communication systems.}

\subsection{DL-Based Wireless Design}

As a pivotal technique of AI, DL models have witnessed an emerging trend of being applied to various aspects of wireless networks, such as power allocation\cite{2017:haoran,2018:haoran}, channel estimation\cite{2018:hengtao}, link scheduling\cite{2024:tingli}, and beamforming design\cite{10032267}. This is attributed to its distinctive  capabilities to learn directly from data without relying on mathematical models and perform fast inference. Regarding the beamforming design, the author of \cite{2020:Junbeom} trained a multi-layer perceptron (MLP) model to optimize power-constrained beamforming vectors for downlink multi-user multi-antenna systems. Additionally, convolutional neural networks (CNNs) were utilized in several studies to learn the mapping between channel state information (CSI) and beamforming design \cite{2020:hao,2020:wenchao,2021:kim}.
Despite  the promising results achieved by these models in specific  scenarios, they suffer from a significant performance degradation with the increasing system complexities, due to their inability to capture the inherent topological structure of wireless networks \cite{2021:yifei}. Moreover, MLP and CNN models, whose input and output dimensionalities are intricately linked to the system scales during the training phase, may struggle to adapt to the dynamic characteristics of wireless networks. 

Recently, driven by the remarkable generalization capabilities of graph neural networks (GNNs) over wireless networks, researchers have shown a growing tendency to adopt GNNs to handle intricate wireless communication problems and achieve wireless intelligence \cite{lu2025agenticgraphneuralnetworks}. For instance, the authors in \cite{2021:tianrui} proposed an unsupervised GNN-based approach, which incorporates a beamforming recovery module to tackle the beamforming design problem in device-to-device (D2D) wireless networks. Similarly, the authors in \cite{2024:yuhang} presented an innovative frequency-efficient beamforming design scheme based on GAT for the  MU-MISO system with per-user power budgets. In addition to homogeneous GNNs, some studies developed intelligent solvers based on heterogeneous GNNs (HGNNs). In particular, Zhang et al. modeled a heterogeneous D2D system in which nodes are categorized based on the number of antennas at both the transmitter and receiver. Building on this model, they proposed a heterogeneous interference GNN (HIGNN) to the joint optimization of power control and beamforming vectors\cite{2021:xiaochen}. However, all of the aforementioned methods were designed from the outset to address specific problems, which limits the model's ability to adapt to different downstream tasks.

\subsection{LLM-Based Wireless Design}

Research related to LLMs in the field of beamforming design is still in its early stages. For instance, the authors in \cite{2025:Sheng} leveraged LLMs to model beamforming prediction as a time series, thereby enhancing the robustness of beamforming. Similarly, the authors in \cite{2025:liu} designed an MoE-LoRA fine-tuning framework, which enables the transfer of LLMs to various downstream tasks, including Radio Environment Modeling, Channel Reconstruction, and Beam Management. Furthermore, the authors in \cite{2025:zheng} proposed BeamLLM to address the high training overhead and latency challenges in the beamforming task for millimeter-wave systems. Building on this, the authors in \cite{2025:tianyue} introduced a multi-task framework based on LLMs, enabling simultaneous multi-user precoding, signal detection, and channel prediction. These advancements primarily focus on the higher-layer communications systems, but similar innovative approaches are beginning to be explored in the context of physical-layer optimization.
For example, the authors in \cite{2024:lee} incorporated channel gains and corresponding transmit power strategies into the LLM and performed power allocation using a few-shot learning approach. The study demonstrated that the LLM can automatically understand and apply the optimal power allocation principle based on the water-filling algorithm without the need for retraining. In \cite{2025:guo}, the authors leveraged LLMs to boost the performance of AI-driven CSI feedback in various contexts. By incorporating the channel distribution as a prompt within the decoder, they significantly improved the accuracy of channel reconstruction and channel prediction\cite{2024:boxun}. The authors in \cite{2024:boxun} froze most of the parameters of the LLM and fine-tuned only a few specific modules to predict future downlink CSI in multiple-input multiple-output systems, thereby enabling efficient cross-modality knowledge transfer. In addition, they designed customized modules tailored to the characteristics of channel data to enhance the model’s adaptability and prediction performance.


\section{Problem Definition and BERT4beam}

\subsection{Problem Definition}
We consider a downlink MU-MISO system, where a transmitter equipped with \(N_{\rm T}\) antennas serves \(K\) single-antenna users over a shared frequency band. Denote the set of users by \(\mathcal{K} \triangleq \{1, 2, \dots, K\}\), where each element represents a user index.


Denote the symbol for the $k$-th user and the corresponding beamforming vector as $s_k$ and ${{{\bf{w}}_k}}\in{\mathbb{C}^{{N_{\rm{T}}}}}$, respectively. The received signal at the $k$-th user is given by
\begin{flalign}
{{\rm{y}}_k} = {\bf{h}}_k^H{{\bf{w}}_k}{s_k} + \sum\nolimits_{i \in {\cal K}\setminus\{k\}} {{\bf{h}}_k^H{{\bf{w}}_i}{s_i}}  + {n_k},
\end{flalign}
where ${{{\bf{h}}_k}}\in{\mathbb{C}^{{N_{\rm{T}}}}}$ denotes the CSI of the link between the transmitter and the $k$-th user, and $n_k\sim\mathcal{CN}( {0,{\sigma_k ^2}})$ denotes the additive white Gaussian noise (AWGN) at the $k$-th user. Without loss of generality, it is assumed that ${{\mathbb E}}\{ {{{| {s_k } |}^2}} \} = 1$ ($\forall k\in\mathcal{K}$). Then, the achievable rate at the $k$-th user is expressed as
\begin{flalign}
{R_k}\left( {\left\{ {{{\bf{w}}_i}} \right\}} \right) = {\log _2}\left( {1 + \frac{{{{\left| {{\bf{h}}_k^H{{\bf{w}}_k}} \right|}^2}}}{{\sum\nolimits_{i \in {\cal K}\setminus\{k\}} {{{\left| {{\bf{h}}_k^H{{\bf{w}}_i}} \right|}^2}}  + {\sigma_k ^2}}}} \right).
\end{flalign} 

Our goal is to maximize the system utility function associated with the achievable rates by finding optimal $\{\mathbf{w}_i\}$ that solve
\begin{subequations}\label{pa}
\begin{align}
&\mathop {\max }\limits_{\left\{{{{\bf{w}}_{i}}}\in{\mathbb{C}^{{N_{\rm{T}}}}} \right\}} { {\rm U}(\{\mathbf{w}_i\}) }  \\
{\rm s.t.}~&{ {\sum\nolimits_{k = 1}^K {\left\| {{{\bf{w}}_{k}}} \right\|_2^2} } } \le {P_{\rm Max}}, \label{power}
\end{align}
\end{subequations}
where ${\rm U}(\{\mathbf{w}_i\})$ denotes the system utility which can be sum rate (SR), i.e., $${\rm U}(\{\mathbf{w}_i\}) = {\sum\nolimits_{k = 1}^K {R_{k}}\left( {\left\{ {{{\bf{w}}_{i}}} \right\}} \right)  }$$ 
or  min rate (MR), i.e., $${\rm U}(\{\mathbf{w}_i\}) = {\min\limits_{k} {R_{k}}\left( {\left\{ {{{\bf{w}}_{i}}} \right\}} \right) }$$ 
or EE, i.e., $${\rm U} \left( {\left\{ {{{\bf{w}}_i}} \right\}} \right) = \frac{{\sum\nolimits_{k = 1}^K {{R_k}\left( {\left\{ {{{\bf{w}}_i}} \right\}} \right)} }}{ \sum\nolimits_{k = 1}^K {\left\| {{{\bf{w}}_k}} \right\|_2^2}  + {P_{\rm C}}},$$ where ${P_{\rm C}}$ denotes the constant power consumption introduced by circuit modules.

\subsection{Concept of BERT4beam}

The aforementioned problems can be addressed by traditional optimization methods. However, these methods often suffer from limited computational efficiency, especially in real-time scenarios. As an alternative, the problem can be reformulated as a DL task. In this case, a (pre-trained) DL model can be trained or fine-tuned to handle a specific task under the ``learning-to-optimize'' paradigm, thereby achieving real-time and near-optimal inference. Nevertheless, wireless networks—characterized by diverse system utilities and large-scale configurations—would require a vast number of task-specific DL models, posing significant challenges for both the development and deployment of such models.

To overcome these limitations, we propose a novel framework termed BERT4beam, which aims to establish a unified large-scale AI model capable of handling a wide range of tasks. For clarity, this paper adopts a triplet $\langle {\rm U}(\{\mathbf{w}_i\}) \in \{\text{EE}, \text{MR}, \text{SR}\}, K, N_{\rm T} \rangle$ to represent a specific beamforming task. This framework is inspired by the perspective of treating beamforming optimization problems as a token-level sequence learning task. The core components of BERT4beam include CSI tokenization, BERT model construction, and dedicated pre-training and fine-tuning strategies.


Based on BERT4beam, we propose two BERT-based approaches for beamforming optimization in the following sections: one tailored for single-task learning and another for multi-task learning.



\section{BERT-Based Single-Task Beamforming Optimization}
This section introduces a BERT-based beamforming approach tailored for single-task beamforming optimization. The  proposed approach comprises three key components, i.e., CSI tokenization, BERT model, and  pre-training and fine-tuning strategies (under the single-task setting).

\subsection{CSI Tokenization}
{For the $i$-th user, the CSI vector $\mathbf{h}_i \in \mathbb{C}^{N_{\rm T}}$ is decomposed into its real and imaginary parts, which are then concatenated to form a single token:}
\begin{equation}\label{tokenization}
    \mathbf{t}_i = \text{Concat}\left(\Re(\mathbf{h}_i)^T, \Im(\mathbf{h}_i)^T\right)^T \in \mathbb{R}^{2N_{\rm T}}.
\end{equation} 
{Notably, the token $\mathbf{t}_i $ generated via \eqref{tokenization} preserves both the amplitude and phase information of $\mathbf{h}_i$, while converting the complex-valued vector $\mathbf{h}_i$ into a real-valued token suitable for processing by BERT.}
Consequently, we denote the sequence of all CSI tokens by a CSI token matrix, i.e., $\mathbf{T} \triangleq \left[ {\mathbf{t}}_1, {\mathbf{t}}_2, \ldots, \mathbf{t}_K \right]^T \in \mathbb{R}^{K\times 2N_{\rm T}}$.

\begin{figure*}[t]
\begin{center}

\includegraphics[ width=1\textwidth]{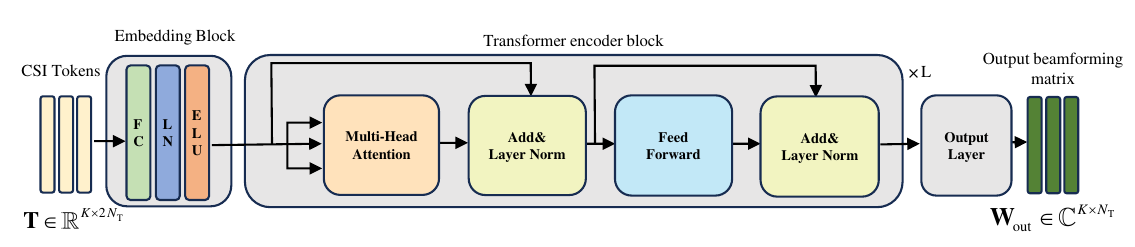}

\captionsetup{justification=justified,singlelinecheck=false}

\caption{{Architecture of the proposed BERT model, which comprises an embedding block, $L$ TEBs, and an output layer. Specifically, the embedding block projects the CSI token matrix $\mathbf{T}$ into a high-dimensional feature space, $L$ TEBs refine the projected features, and the output layer maps the refined features to the feasible beamforming matrix $\mathbf{W}_{\text{out}}$.}}

\label{BERT:architecture}
\end{center}
\end{figure*}
\subsection{BERT Model}
For a specific task, the CSI token is fed into a task-tailored BERT model to generate beamforming vectors. As illustrated in Fig. \ref{BERT:architecture}, the proposed BERT model consists of three core modules, i.e., an embedding block, $L$ transformer encoder blocks (TEBs), and an output layer. The operations and implementation details of each module are described as follows.


\subsubsection{Embedding Block}
The embedding block projects the input CSI token into a high-dimensional feature space aligned with the input dimension of the subsequent TEB. 
The embedding block is realized by a fully-connected (FC) layer followed by layer normalization (LN) and nonlinear activation function.  

%


The LN stabilizes the learning process and mitigates the issue of internal covariate shift\cite{layernorma}. The mathematical expression for LN is given by

{
\begin{equation}
\mathrm{LN}(\mathbf{x}) = \frac{\mathbf{x} - \mu \mathbf{1}}{\sigma} \cdot \gamma + \beta \mathbf{1},
\end{equation}}
where \(\mathbf{x}\) is the input vector, \(\mu\) and \(\sigma\) are the mean and standard deviation of the input across the feature dimension, respectively, \(\gamma\) and \(\beta\) are learnable parameters that scale and shift the normalized output, {and $\mathbf{1}$ denotes an all-ones vector with the same dimension as $\mathbf{x}$.}

We employ the exponential linear unit (ELU) as the activation function to introduce nonlinearity as  well as enhance the model's ability to learn complex patterns\cite{ELU}. The mathematical expression for the ELU function is given by
\begin{equation}
\text{ELU}(x) =
\begin{cases}
x, & \text{if } x > 0, \\
\alpha (\exp(x) - 1), & \text{if } x \leq 0,
\end{cases}
\end{equation}
where \(x\) is the input of the ELU function, and \(\alpha\) is a constant  (usually set to 1)  that controls the value for negative inputs.

The embedding block is expressed as
\begin{equation}
    \mathbf{T}_{\text{emb}} = \text{ELU}\left(\text{LN}\left( \mathbf{T}\mathbf{W}_\text{fc} \right)\right) \in \mathbb{R}^{K \times F},
\end{equation}
where $\mathbf{T}_{\text{emb}}$ denotes the updated $\mathbf{T}$, \(\mathbf{W}_\text{fc} \in \mathbb{R}^{2N_{\rm T} \times F}\)  represents the weights of the FC layer, and $F$ represents the output embedding dimension. 

\subsubsection{Transformer Encoder Block}
The TEBs play a pivotal role in learn the beamforming vectors. Each TEB primarily comprises  two sub-blocks: the  multi-head attention sub-block and the   position-wise FC feed-forward sub-block. Each sub-block incorporates a residual connection and an LN operation to facilitate the training process  and prevent the vanishing gradient problem. In the following, one of the $L$ TEBs is selected as an example to illustrate the processes of the two sub-blocks in detail; the other TEBs can be explained in a similar way.

{\it Sub-block $1$: Multi-head attention.} The first sub-block leverages the bidirectional multi-head attention (MHA) mechanism to compute attention weights (cf. \eqref{att}) among the CSI tokens, aiming to capture contextual dependencies. Then, the obtained attention weights are utilized to derive deep bidirectional representations (cf. \eqref{dbr}) via weighted summation, which are further used for downstream beamforming design tasks.

The MHA mechanism adopts the scaled dot-product attention mechanism. Denote  $C$  by the number  of attention heads of the MHA mechanism. Given $\mathbf{T}_{\text{emb}}$, the MHA mechanism computes the output of the $c$-th head as
\begin{equation}\label{att}
    \mathbf{O}^{(c)} = \text{Softmax} \left( \frac{\mathbf{Q}^{(c)} \left({\mathbf{K}^{(c)}}\right)^T}{\sqrt{d}} \right) \mathbf{V}^{(c)},
\end{equation}
where \(d = F / C\) is the dimension of each head, and the query, key, and value matrices $\mathbf{Q}^{(c)},\mathbf{K}^{(c)},\mathbf{V}^{(c)} \in \mathbb{R}^{K \times F}$ are obtained by linear projections, i.e.,
\[
\mathbf{Q}^{(c)} = \mathbf{T}_{\text{emb}} \mathbf{W}_Q^{(c)},~
\mathbf{K}^{(c)} = \mathbf{T}_{\text{emb}}  \mathbf{W}_K^{(c)},~
\mathbf{V}^{(c)} = \mathbf{T}_{\text{emb}}  \mathbf{W}_V^{(c)},
\]
where \(\mathbf{W}_Q^{(c)}, \mathbf{W}_K^{(c)}, \mathbf{W}_V^{(c)} \in \mathbb{R}^{F \times d}\) are learnable projection matrices for the \(c\)-th head. Then, the MHA mechanism concatenates \(C\) attention heads followed by a linear transformation to obtain its output as
\begin{equation}\label{dbr}
 \text{MHA}(\mathbf{T}_{\text{emb}}) = \text{Concat}(\mathbf{O}^{(1)}, \mathbf{O}^{(2)}, \ldots, \mathbf{O}^{(C)}) \mathbf{W}_O,   
\end{equation}
where \(\mathbf{W}_O \in \mathbb{R}^{F \times F}\) is the learnable linear transformation matrix, and  \( \text{Concat}(\cdot) \) denotes the concatenation operation.

With the integration of residual connection and the LN operation, the first sub-block is expressed as
\begin{equation}
\mathbf{T}_{\text{fir}} = \text{LN}\left( \mathbf{T}_{\text{emb}} + \text{MHA}(\mathbf{T}_{\text{emb}}) \right) \in \mathbb{R}^{K \times {F}},    
\end{equation}
where $\mathbf{T}_{\text{fir}}$ is the output token sequence of the first sub-block.

{\it Sub-block $2$: Position-wise fully-connected feed-forward.} 

The core of the second sub-block is the feed-forward neural network (FFN), which is composed of two FC layers, with a nonlinear activation function interposed between them. Mathematically, the FFN is expressed as

{
\begin{equation}
\mathbf{FFN}(\mathbf{T}_{\text{fir}}) = 
\text{GELU}\Big( \mathbf{T}_{\text{fir}} \mathbf{W}_{\text{ffn}} + \mathbf{1}_{\text{ffn}} \mathbf{b}_{\text{ffn}}^T \Big) \widehat{\mathbf{W}}_{\text{ffn}} + \mathbf{1}_{\text{ffn}} \widehat{\mathbf{b}}_{\text{ffn}}^T,
\end{equation}}

where \(\mathbf{W}_{\text{ffn}} \in \mathbb{R}^{F \times d'}\)/\(\mathbf{b}_{\text{ffn}} \in \mathbb{R}^{d'}\)  and \(\widehat{\mathbf{W}}_{\text{ffn}} \in \mathbb{R}^{d' \times F}\)/\( \widehat{\mathbf{b}}_{\text{ffn}} \in \mathbb{R}^{F}\)  denote the learnable  matrices/bias vectors associated with the two FC layers, respectively; \(d'\) is the dimension of the hidden layer, {and $\mathbf{1}_{\text{ffn}} \in \mathbb{R}^{K}$ denotes an all-ones vector.} We adopt the gaussian error linear unit (GELU) as the activation function, which is given by
\begin{equation}
\text{GELU}(x) = 0.5 x \left( 1 + \tanh\left( \sqrt{\frac{2}{\pi}} (x + 0.044715 x^3) \right) \right),
\end{equation}
where \(x\) is the input to the GELU function, and \(\tanh(\cdot)\) represents the hyperbolic tangent  function.

By combining with the residual connection and the LN operation, the second sub-block is expressed as 
\begin{equation}
   \mathbf{T}_{\text{sec}} = \text{LN}\left( \mathbf{T}_{\text{fir}} + \text{FFN}(\mathbf{T}_{\text{fir}}) \right) \in \mathbb{R}^{K \times F},  
\end{equation}
where $\mathbf{T}_{\text{sec}}$ is the output token sequence of the second sub-block.

\subsubsection{Output Layer}

The output layer consists of an FC layer and a generalizable power adapter (GPA).

The FC layer maps \(\mathbf{T}_{\text{sec}}\), produced by the last TEB, into the target dimension, i.e., \(N_{\rm T}\).  Denote $\mathbf{W}_{\text{out}} \in \mathbb{C}^{K \times N_{\rm T}}$ as the complex-valued beamforming matrix. The FC layer is formulated as
\begin{equation}
    \mathbf{W}_{\text{out}} = \mathbf{T}_{\text{sec}}{\mathbf{W}}_{\text{real}} + j  \mathbf{T}_{\text{sec}} {\mathbf{W}}_{\text{imag}},
\end{equation}
where  \(\mathbf{W}_{\text{real}} \in \mathbb{R}^{F \times N_{\rm T}}\) and \(\mathbf{W}_{\text{imag}} \in \mathbb{R}^{F \times N_{\rm T}}\) are the learnable parameters.

The GPA is applied to $\mathbf{W}_{\text{out}}$ to enforce the total transmit power constraint (\ref{power}). Mathematically, the GPA is expressed as 
{
\begin{flalign} 
\mathbf{W}_{\text{out}} :&=\label{act}
{\rm GPA}\left({\mathbf{W}}_{\text{out}} \right) \\&= \left\{
\begin{array}{ll}
\sqrt{P_{\rm Max}}  {\mathbf{W}}_{\text{out}}, & \quad \|{\mathbf{W}}_{\text{out}}\|^2 \le 1, \\
\sqrt{\frac{P_{\rm Max}}{\|\mathbf{W}_{\text{out}}\|}}  {\mathbf{W}}_{\text{out}}, & \quad \| {\mathbf{W}}_{\text{out}}\|^2 > 1.
\end{array}
\right. \nonumber
\end{flalign}}

Notably, the GPA is parameter-free, which allows the model to effectively adapt to varying power budget constraints.

The output layer is expressed as 
\begin{flalign}
    \mathbf{W}_{\text{out}} &\triangleq \left[{\bf w}_1,{\bf w}_2,\ldots,{\bf w}_K\right]\\
    &={\rm GPA}\left(\mathbf{T}_{\text{sec}}{\mathbf{W}}_{\text{real}} + j  \mathbf{T}_{\text{sec}} {\mathbf{W}}_{\text{imag}}\right).\nonumber
\end{flalign}

\subsection{Pre-Training and Fine-Tuning Strategies}

The training process consists of two stages, i.e., the pre-training stage and the fine-tuning stage. The first stage is based on supervised pre-training on a large labeled dataset to obtain a pre-trained model for a specific task. The second stage further refines the model through unsupervised fine-tuning, with simple modifications applied to the pre-trained model. 

\subsubsection{Supervised Pre-Training}


For a specific downstream task, we first generate the labeled  beamforming matrix denoted by \( \mathbf{W}_{\text{cvx}}\in \mathbb{C}^{K \times N_{\rm T}} \) via traditional optimization algorithms.  Then, we adopt the supervised loss function that consists of two parts: one is the cosine similarity between \( \mathbf{{W}}_{\text{cvx}} \) and  \( \mathbf{{W}}_{\text{out}} \), and the other is the system utility for the specific task. The loss function is formulated as 
\begin{equation}\label{fun:loss}
  \mathcal{L}_{\text{pre}} = \lambda_1 \cdot \left( 1 - \frac{\left| \mathbf{W}_{\text{cvx}}^H \mathbf{W}_{\text{out}} \right|^2}{\|\mathbf{W}_{\text{cvx}}\|^2 \|\mathbf{W}_{\text{out}}\|^2} \right) - \lambda_2 \cdot {\rm U}(\mathbf{W}_{\text{out}}),    
\end{equation}
where \( \lambda_1 \in \mathbb{R} \) and \( \lambda_2 \in \mathbb{R} \) are hyperparameters that balance the cosine similarity loss and the task utility loss.


\begin{remark} (Generalization to varying user scales.)
Owing to the parameter-sharing mechanism of the BERT model, its input and output dimensions are independent of the number of users, i.e., $K$, thereby enabling direct generalization to tasks with varying user scales (without the need for fine-tuning). 
\end{remark}

\begin{remark} (Adaptation to varying system utilities and  antenna configurations.) 
Empowered by the MHA mechanism, the BERT model is capable of capturing the underlying representations of diverse tasks during pre-training. That is, the parameters of TEBs are well initialized.  As a result, when deployed in tasks with different system utilities or/and number of antennas, the BERT model can rapidly adapt and deliver efficient beamforming solutions via fine-tuning.
\end{remark}


\subsubsection{Unsupervised Fine-Tuning}
Fine-tuning refers to the process of training a pre-trained model by utilizing the parameters that it has already learned, instead of initializing those parameters randomly. Particularly, for tasks with a different number of antennas from the pre-training task, we load the parameters of pre-trained TEBs while reconfiguring the embedding block and output layer to match the number of antennas and randomly initializing them. For other tasks, we can load all parameters of the pre-trained BERT model.

Compared to pre-training, fine-tuning is computationally efficient, requiring only a small number of unlabeled samples and a few epochs. The unsupervised loss function is expressed as
\begin{equation}\label{Fine-tuning-loss}
  \mathcal{L}_{\text{fin}} = - {\rm U}(\mathbf{\mathbf{W}_{\text{out}}}).
\end{equation}

\section{UBERT-Based Multi-Task Beamforming Optimization}

To enhance the model's cross-task generalization and adaptive capabilities, this section presents a unified BERT-based approach for multi-task beamforming design, termed  UBERT. Specifically, the proposed approach incorporates three key components, i.e., element-wise tokenization, UBERT mode, and pre-training and fine-tuning strategies (under the multi-task setting).

\subsection{Element-Wise Tokenization}

UBERT adopts a finer-grained tokenization strategy, i.e., element-wise tokenization. Specifically, we partition $\mathbf{h}_i$ into \( N_{\rm T} \) tokens. Thus, we obtain \( K \times N_{\rm T} \)  tokens for the considered system with \( K \) users. For convenience, the sequence  tokens are organized into an antenna token matrix denoted by \( \mathbf{T}_{\text{a}} \in \mathbb{R}^{K \times N_{\rm T} \times 2} \), where each \( \mathbf{T}_{\text{a}}[i,j] \) consists of the concatenation of the real and imaginary parts, and is given by
\begin{equation}
\mathbf{T}_{\text{a}}[i, j] =  \text{Concat}\left( \Re(\mathbf{h}_{i}[j]), \Im(\mathbf{h}_{i}[j] \right)  \in \mathbb{R}^{2}.
\end{equation}
Notably, the element-wise tokenization guarantees that the model's input dimensions remain independent of the system scales, including the number of users \( K \) and antennas \( N_{\rm T} \).

\begin{figure*}[t]
\begin{center}
{
\includegraphics[ width=1\textwidth]{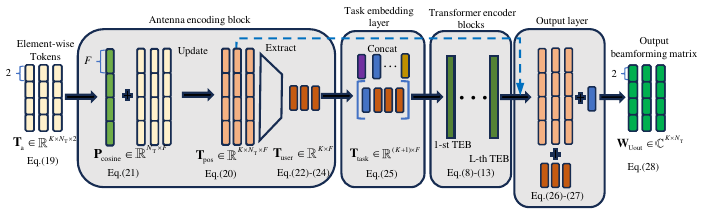}
}
\captionsetup{justification=justified,singlelinecheck=false}
\caption{{Architecture of the proposed UBERT model, which comprises an antenna encoding block, a task embedding layer, $L$ TEBs, and an output layer. The antenna encoding block extracts user tokens from the antenna token matrix, the task embedding layer incorporates task awareness, $L$ TEBs refine the tokens, and the output layer generates the feasible beamforming matrix $\mathbf{W}_{\text{Uout}}$.}}
\label{UBERT}
\end{center}
\end{figure*}

\subsection{UBERT Model}
The overall architecture of UBERT is illustrated in Fig. \ref{UBERT}, which consists of an antenna encoding block (AEB), a task embedding layer, $L$ TEBs, and an output layer.

\subsubsection{Antenna Encoding Block}
The AEB captures the correlations among antenna tokens via an attention mechanism, and generates user tokens (cf. \eqref{ut}) through an extraction operation. The AEB is composed of three main modules: a  positional encoding layer, an attention-based token update layer, and a user token extraction layer.

The positional encoding  layer first projects antenna tokens associated with the $k$-th user, i.e., ${\bf T}_{\rm a}[k]$, into higher-dimensional representations (equal to the dimension \( F \) of the TEB), and then incorporates them and the relative positional information of antennas via cosine positional encoding.  The positional encoding layer is expressed as
\begin{equation}
    \mathbf{T}_{\text{pos}}{[k]} = \mathbf{T}_{\text{a}}{[k]} \cdot \mathbf{W}_\text{embedding} + \mathbf{P}_\text{cosine}\in \mathbb{R}^{N_{\rm T} \times F},
\end{equation}
where {$\mathbf{T}_{\text{pos}}{[k]}$ denotes the updated token matrix of $\mathbf{T}_{\text{a}}{[k]}$,} \( \mathbf{W}_\text{embedding} \in \mathbb{R}^{2 \times F} \) denotes the embedding weight matrix, and \( \mathbf{P}_\text{cosine}  \in \mathbb{R}^{N_{\rm T} \times F}\)  denotes the cosine positional encoding matrix, where the $\langle i,j \rangle$-th element is calculated using the following formula
\begin{equation}
\mathbf{P}_\text{cosine}[i, j] = \begin{cases}
\cos\left(\frac{i}{10000^{2j/F}}\right), & \text{for even } j, \\
\sin\left(\frac{i}{10000^{2j/F}}\right), & \text{for odd } j,
\end{cases}
\end{equation}
where \( i \) represents the antenna index (\( i \in\{ 1, 2, \dots, N_{\rm T}\} \)) and \( j \) represents the index of the embedding dimension (\( j \in\{ 1, 2, \dots, F \}\)).   Notably, this layer can enhance the attention mechanism’s capability to model the relationships among tokens  and improve the model's sensitivity to variations in antenna configurations\cite{Attention_Illia}.

The attention-based token update layer employs  an  additive attention mechanism, where the inter-element attention weights for   $\mathbf{T}_{\text{pos}}$ are computed. Let \( \mathbf{A} \in \mathbb{R}^{K \times N_{\rm T} \times N_{\rm T}} \) denote the attention coefficient matrix, which is calculated by
\begin{equation}
\mathbf{A}{[k,i,j]} = {{\rm ReLU}\left(\mathbf{a}^{T} \mathbf{W}_s\mathbf{T}_{\text{pos}}[k,i]
+ \mathbf{a}^{T} \mathbf{W}_t\mathbf{T}_{\text{pos}}[k,j]  \right),}
\end{equation}
where {\( \mathbf{W}_s, \mathbf{W}_t \in \mathbb{R}^{F \times F} \)} are the attention weight matrices for the source and target tokens, respectively, and $\mathbf{a} \in \mathbb{R}^{F}$ represents the learnable attention vector. Then, $\mathbf{T}_{\text{pos}}[k]$ is updated based on $\mathbf{A}{[k]}$, allowing each token to capture global antenna features. The attention-based token update layer for the \( i \)-th token of the \( k \)-th user is given by 
\begin{equation}\label{equ:ant_token}
    {\mathbf{T}_{\text{ant}}}[k,i] = \mathrm{Softmax}\left(\mathbf{A}[k,i]\right)^{T}\mathbf{T}_{\text{pos}}[k]\in \mathbb{R}^{F},
\end{equation}
{where ${\mathbf{T}_{\text{ant}}}\in\mathbb{R}^{K \times N_{\rm T} \times F}$  denotes the updated token matrix of $\mathbf{T}_{\text{pos}}$.}

The  user token extraction layer aggregates $\mathbf{T}_{\rm ant}$ for each user to extract the user token matrix, denoted by ${\mathbf{T}_{\text{user}}}\in\mathbb{R}^{K \times F}$, via a summation operation (i.e., ${\rm SUM}(\cdot)$). The $k$-th user token is expressed as   
\begin{equation}\label{ut}
\mathbf{T}_{\text{user}}[k] = \mathbf{W}_{\text{ext}} \cdot \text{SUM}\left(\mathbf{T}_{\text{ant}}[k]\right) \in \mathbb{R}^{F},
\end{equation}
where \(\mathbf{W}_{\text{ext}} \in \mathbb{R}^{F \times F}\) is a learnable linear transformation matrix.
\subsubsection{{Task Embedding Layer}} 
The task embedding layer is designed to endow UBERT with the capability to distinguish  diverse beamforming tasks. Denote \( \mathbf{t}_{\text{task}}\in {\mathbb R}^F \) as the task embedding token, which is trainable. Then, the task embedding layer concatenates the task token $\mathbf{t}_{\text{task}}$ with the user token matrix $\mathbf{T}_{\text{user}}$:
\begin{equation}
    \mathbf{T}_{\text{task}} = \text{Concat}(\mathbf{T}_{\text{user}},\, \mathbf{t}_{\text{task}}) \in \mathbb{R}^{(K+1)\times F},
\end{equation}
where $\mathbf{T}_{\text{task}}$ denotes the task-aware user token matrix, {and $\text{task} \in \{\text{EE}, \text{SR}, \text{MR}\}$ represents the specific optimization objective.}

\subsubsection{Transformer Encoder Block} 

The TEB of UBERT is consistent with the TEB introduced in section III-B. Let \( \mathbf{T}_{\text{teb}} \in \mathbb{R}^{(K+1)\times F}\) represent the token sequence output by the last TEB, where $\mathbf{T}_{\text{ted}}[K+1]$ represents the updated task embedding. That is, $L$ TEBs map $\mathbf{T}_{\text{task}}$ to $\mathbf{T}_{\text{teb}}$.

\subsubsection{{Output Layer}} 

We feed  $\mathbf{T}_{\text{teb}}$ and $\mathbf{T}_{\text{ant}}$ (cf. \eqref{equ:ant_token}) into the output layer. 

First, $\mathbf{T}_{\text{ted}}[K+1]$ is added to all the updated user tokens, i.e., $\{\mathbf{T}_{\text{ted}}[k]\}$, expressed as follows:
\begin{equation}
     \widehat{\mathbf{T}}_{\text{task}}[k] = \mathbf{T}_{\text{teb}}[k] + \mathbf{T}_{\text{teb}}[K+1] \in \mathbb{R}^{F},
\end{equation}
where $\widehat{\mathbf{T}}_{\text{task}} \in \mathbb{R}^{K \times F}$  represents the task-integrated user token matrix.

By combining   \( \widehat{\mathbf{T}}_{\text{task}} \) with $\mathbf{T}_{\text{ant}}$, we can obtain a new antenna token matrix, denoted by $\widehat{\mathbf{T}}_{\text{ant}}$, that incorporates all the relevant user information, as shown in the following equation:
\begin{equation}
    \widehat{\mathbf{T}}_{\text{ant}}[k,i] = \mathbf{T}_{\text{ant}}[k,i] + \widehat{\mathbf{T}}_{\text{task}}[k]\in{\mathbb R}^F.
\end{equation}

Finally, the output beamforming matrix of the UBERT model, denoted by \( \mathbf{W}_{\text{Uout}} \in \mathbb{C}^{K \times N_{\rm T}} \), is obtained by
\begin{equation}
    \mathbf{W}_{\text{Uout}}[i,j] = {\rm GPA}\left({\widehat{\mathbf{T}}_{\text{ant}}}[i,j]\mathbf{w}_{\text{Urel}} + j {\widehat{\mathbf{T}}_{\text{ant}}}[i,j]\mathbf{w}_{\text{Uimg}}\right),
\end{equation}
where \( \mathbf{w}_{\text{Urel}},\mathbf{w}_{\text{Uimg}} \in \mathbb{R}^{F} \) represent the learnable mapping vectors.

\subsection{Pre-Training and Fine-Tuning Strategies}

The training process of UBERT also consists of pre-training and fine-tuning stages. However, to ensure that UBERT acquires the ability to handle multiple tasks, both stages are supervised.
In the pre-training stage, UBERT learns universal beamforming features from a large-scale labeled dataset,  yielding a well-initialized pre-trained model for different tasks. 
In the fine-tuning stage, UBERT is further optimized to adapt to specific tasks (which may differ from the task during the pre-training stage). Notably, UBERT does not require any architectural modification during fine-tuning thanks to the element-wise tokenization, thus preventing the loss of pre-trained weights. This enables UBERT to maintain high performance with a small number of fine-tuning samples.

\subsubsection{Supervised Pre-Training}

To enable UBERT to learn generalizable knowledge across tasks and become a universal pre-trained model, we design a supervised pre-training loss function composed of multiple task-specific loss components. This can be expressed as:
\begin{equation}\label{UBert_pre}
    \mathcal{L}_{\text{u-pre}} = \mathcal{L}_{\text{EE}} + \mathcal{L}_{\text{SR}}  + \mathcal{L}_{\text{MR}},
\end{equation}
where \(\mathcal{L}_{\text{EE}}\), \(\mathcal{L}_{\text{SR}}\), and \(\mathcal{L}_{\text{MR}}\) represent the loss functions corresponding to EE, SR, and MR, respectively.

To avoid potential issues such as gradient conflicts or large discrepancies in gradient magnitudes during the gradient descent process, the loss function for each task is specifically formulated as
\begin{equation}
\mathcal{L} = \frac{1}{K \cdot N_{\rm T}} \left( 1 - \frac{1}{K} \text{Trace} \left( {\bf W}_{\text{Uout}}^{\text{full}} \cdot \left({\bf W}_{\text{Ucvx}}^{\text{full}}\right)^T \right) \right),
\end{equation}
where {$\mathcal{L}\in\{\mathcal{L}_{\text{EE}}, \mathcal{L}_{\text{SR}}, \mathcal{L}_{\text{MR}}\}$} and
$$\mathbf{W}_{\text{Uout}}^{\text{full}} =  \text{Concat}\left(\Re(\mathbf{W}_{\text{Uout}}), \Im(\mathbf{W}_{\text{Uout}})\right),$$
$$\mathbf{W}_{\text{Ucvx}}^{\text{full}} =  \text{Concat}\left(\Re(\mathbf{W}_{\text{Ucvx}}), \Im(\mathbf{W}_{\text{Ucvx}})\right),$$
where \(\mathbf{W}_{\text{Ucvx}} \in \mathbb{C}^{K \times N_{\rm T}}\) represents the labels obtained from traditional optimization algorithms.

Furthermore, to ensure a smooth pre-training process, we design a multi-task training algorithm, which is summarized in Algorithm \ref{alg}.  Each batch uniformly samples tasks from all available tasks to ensure balanced optimization across tasks during training.

\subsubsection{Supervised Fine-Tuning}
UBERT employs the identical model architecture and loss function for fine-tuning as those used in the pre-training phase.

\begin{remark} (Generalization to diverse tasks.)
During the pre-training phase, UBERT can learn universal beamforming design features, enabling it to adapt to various beamforming tasks and capture task-shared features. Additionally, UBERT’s input and output dimensions are completely independent of the system scale, and thus, its architecture can remain unchanged between the pre-training and fine-tuning stages. This architectural consistency helps to fully preserve the knowledge acquired during pre-training, thereby enabling more efficient and effective generalization to diverse tasks in the fine-tuning stage.
\end{remark}


\begin{algorithm}[h]
\caption{{Multi-task Pre-Training Algorithm with Uniform Sampling}}
\label{alg}
\begin{algorithmic}[1]
\STATE {\textbf{Input:} Multi-task dataset $\{\mathcal{D}_{\text{EE}}, \mathcal{D}_{\text{SR}}, \mathcal{D}_{\text{MR}}\}$, maximum training iterations $T$, batch size $B$, and multi-task set $\mathcal{T} = \{\text{EE}, \text{SR}, \text{MR}\}$;}
\STATE {\textbf{Output:} Trained model parameters $\mathbf{\Theta}$;}
\STATE {Initialize model parameters $\mathbf{\Theta}$ randomly};
\FOR{$t = 1 \to T$}
    \STATE \texttt{//Processing per epoch}
    \STATE {Initialize an empty batch container with capacity $B$};
    \STATE {Calculate samples per task via uniform sampling: $B_{\text{task}} = B / |\mathcal{T}|$};
    \FOR{each task $\tau \in \mathcal{T}$} 
    \STATE \texttt{//construct batch}
        \STATE {Randomly sample $B_{\text{task}}$ samples from $\mathcal{D}_{\tau}$};
        \STATE {Add the sampled $B_{\text{task}}$ samples to the batch container}; 
    \ENDFOR
    \STATE \texttt{//Update model parameters}
    \STATE {Compute loss $\mathcal{L}_{\text{u-pre}}$  based on the constructed batch and current model parameters $\mathbf{\Theta}$};
    \STATE {Update model parameters $\mathbf{\Theta}$ using the gradient descent optimizer;};
\ENDFOR
\end{algorithmic}
\end{algorithm}

\section{Numerical Results}
This section presents extensive numerical results to validate the effectiveness of the proposed method. 

\subsection{Experimental Setup}


\subsubsection{Simulation Scenario}  The number of antennas $N_{\rm T}$  is varied over a set  of $\{11, 12, 13, 15, 16, 17\}$. The number of users $K$ is selected from the set $\{5, 6, 7, 8, 9\}$. The  power budget is set to $P_{\rm Max} \in \{1, 2, 3\}$ W, and the constant power $P_{\rm C}$ is set to $0.5$ W. The Rayleigh fading channel model is  adopted with the average signal-to-noise-ratio of $10$ dB.

\subsubsection{Dataset}
{The dataset is divided into a pre-training set and a fine-tuning set. The pre-training set includes two system configurations, namely $(K=6, N_{\rm T}=12)$ and $(K=8, N_{\rm T}=16)$, containing $80,000$ labeled samples. The fine-tuning set also includes two configurations, namely $(K \in \{5, 7\}, N_{\rm T} \in \{11, 13\})$ and $(K \in \{7, 9\}, N_{\rm T} \in \{15, 17\})$, with $10,000$ labeled samples. Both the pre-training and fine-tuning sets are randomly split into training, validation, and test sets at a ratio of $8:1:1$.
Each labeled sample consists of a pair $(\mathbf{H}, \mathbf{W})$, where the channel matrix $\mathbf{H}$ is generated based on a Rayleigh fading channel model, i.e.,
\begin{equation}
{\mathbf{H}[i,j] \sim \mathcal{CN}(0,1), \quad \forall i \in \mathcal{K}, \forall j \in \mathcal{N_{\rm T}}.}
\end{equation}
The corresponding beamforming matrix $\mathbf{W}$ is obtained using a traditional CVX optimization algorithm with a convergence accuracy of $10^{-4}$.}


\subsubsection{Implementation Detail} 

The pre-training learning rate is set to $2 \times 10^{-4}$ with a cosine decay schedule. The model is trained for 100 epochs on the pre-training dataset, and the best model weights on the validation set are retained. Fine-tuning\footnote{A large mismatch in the number of users between deployment and training may degrade model performance, while changes in antenna configurations can cause input–output dimensional mismatches for the pre-trained model.} is primarily conducted in scenarios where the number of users $K$ varies significantly or the number of antennas $N_{\rm T}$ changes. During fine-tuning, full-parameter updating is adopted with an initial learning rate of $2 \times 10^{-5}$, and the process converges within only 10 epochs. All training processes use the Adam optimizer with a batch size of 32. The experimental environment consists of PyTorch 2.1.2, Python 3.10 (running on Ubuntu 22.04), and CUDA 11.8. CVX formulation is solved using the CVX solver \texttt{SeDuMi} under MathWorks MATLAB R2021b. All experiments are carried out on a server equipped with a H20-NVLink GPU (96GB), an AMD EPYC 9K84 CPU (96 cores), and 150GB of memory.

\subsubsection{Model Architecture} 
The proposed BERT-based models adopt an embedding dimension of $1,024$, with $12$ TEBs and $16$ attention heads.

\subsubsection{Performance Metric} 
The performance of the problems solved by the proposed models is measured using the following metric over the test set:
\begin{equation}\label{pm}
    {\text{Performance}} = \frac{\frac{1}{N_{{\text{test}}}} \sum_{n=1}^{N_{{\text{test}}}} {\rm U}_{{\text{NN}}}^{(n)}}{\frac{1}{N_{{\text{test}}}} \sum_{n=1}^{N_{{\text{test}}}} {\rm U}_{{\text cvx}}^{(n)}} \times 100\%,
\end{equation}
where \({\rm U}_{{\text{NN}}}^{(n)} \) and \({\rm U}_{{\text{cvx}}}^{(n)} \) represent the performance values of the neural network and the CVX solver for the \(n\)-th sample, respectively, and \( N_{{\text{test}}} \) denotes the number of samples in the test set.

\subsubsection{Baseline Methods} The following baseline methods are considered for performance comparison:
\begin{itemize}
\item \textbf{Successive Convex Approximation (SCA)} \cite{2020:Walid}: This approach is an SCA-based algorithm, implemented using the CVX toolbox for efficient computation. Note that SCA also serves as the method to generate  \( {\rm U}_{{\text{cvx}}}^{(n)} \) in \eqref{pm}.
\item \textbf{MLP} \cite{2020:Junbeom}: This method leverages an FC deep neural network that processes concatenated channel samples as input vectors for the network.
\item \textbf{CNN} \cite{2021:kim}: A CNN is employed to extract feature representations from the channel matrix, which are then utilized for downstream beamforming design tasks.
\item \textbf{GCN} \cite{2021:tianrui}: This approach utilizes a basic graph convolutional network (GCN) model that implements a message-passing mechanism through graph convolution operations.
\item \textbf{GAT} \cite{2024:yuhang}: The model extends the GCN framework by incorporating an MHA mechanism based on additive attention, thereby enhancing its ability to capture complex relationships in graph data.
\item \textbf{GPT} \cite{2025:zheng}: GPT is based on the transformer architecture, utilizing a pure decoder structure. In contrast to BERT, GPT employs a unidirectional attention mechanism.
\end{itemize}

\begin{figure}
\centering 
\subfigure[SR maximization]{\label{fig:a}\includegraphics[width=1\linewidth]{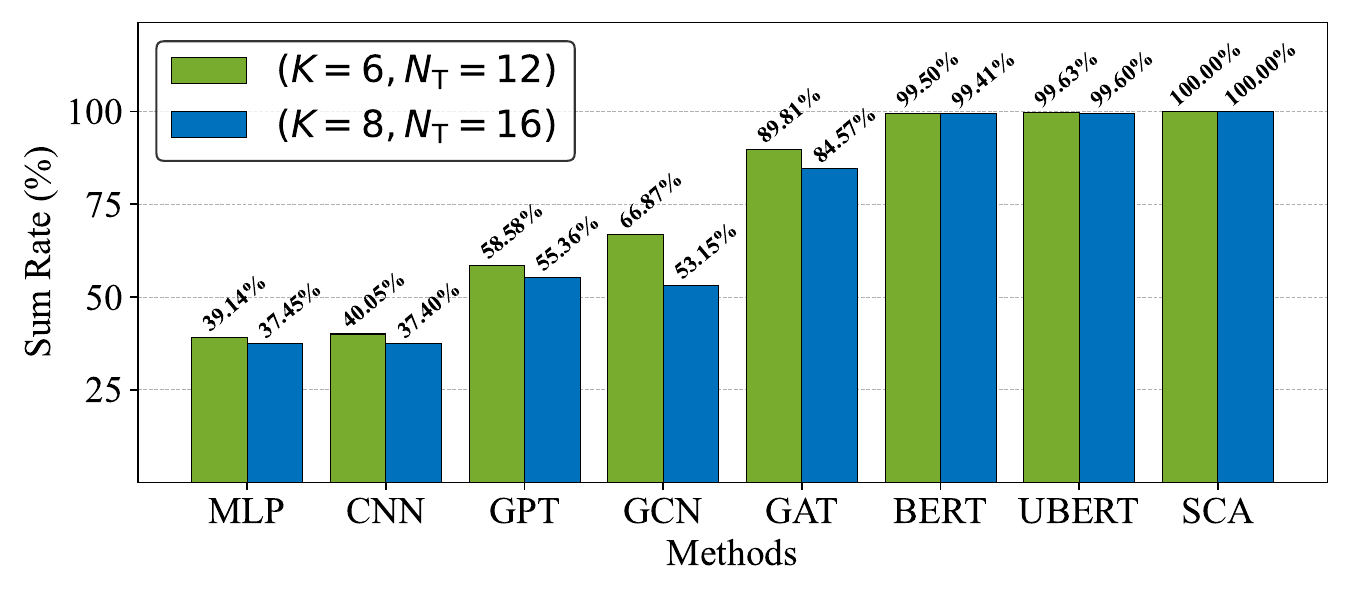}}
\subfigure[MR maximization]{\label{fig:b}\includegraphics[width=1\linewidth]{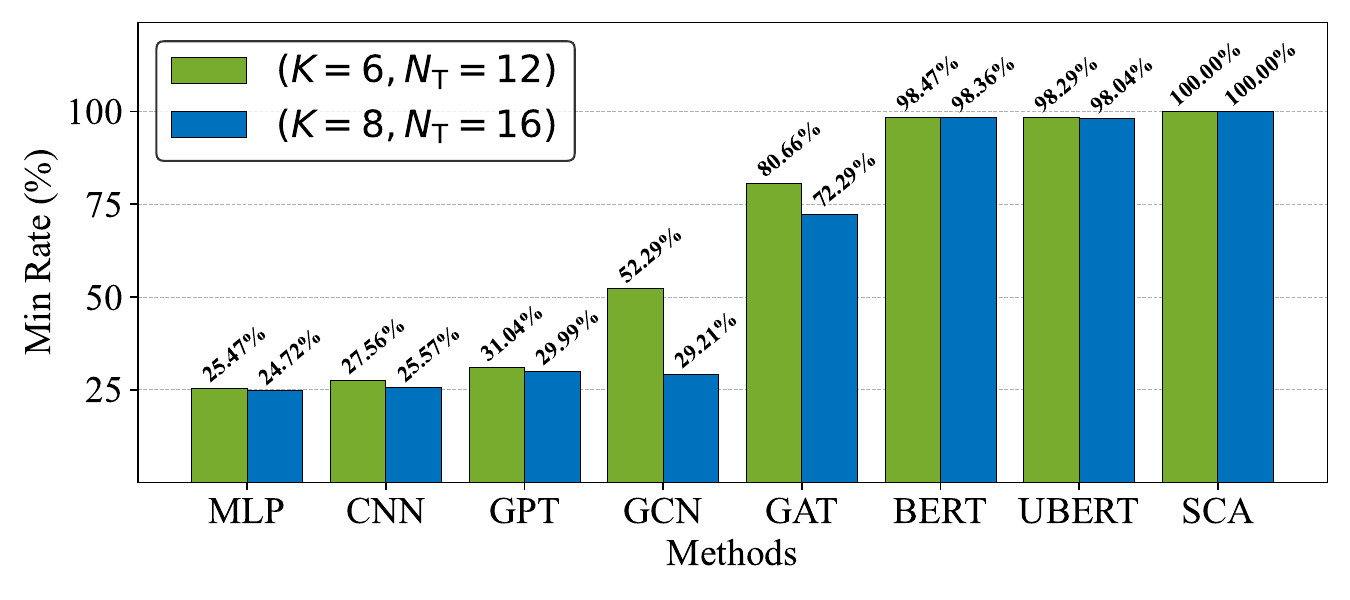}}
\subfigure[{EE maximization}]{\label{fig:c}\includegraphics[width=1\linewidth]{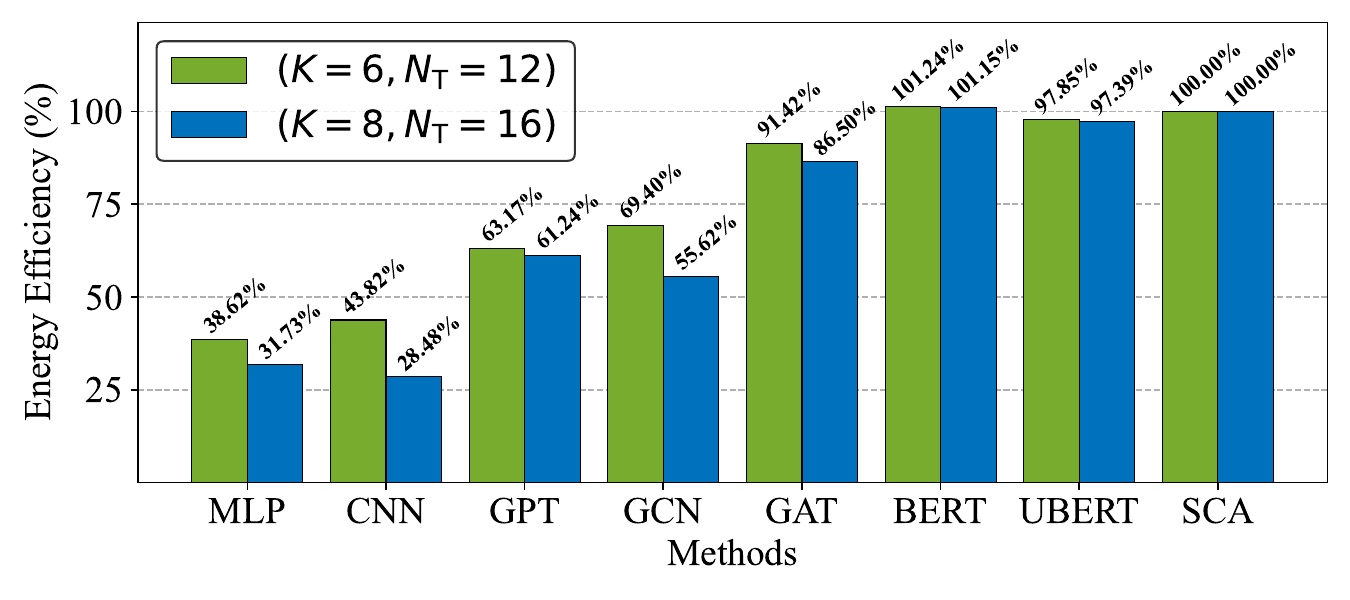}}
\caption{Performance across different system utilities.}
\label{fig:re:pertrain}
\end{figure}

\subsection{Effectiveness of Pre-Trained Models}

This subsection aims to evaluate the effectiveness of the pre-trained models. Specifically, we assess the pre-training performance of each model and analyze its adaptability under different power budgets. Additionally, we test the generalization capability of the pre-trained models with respect to variations in both the power budget and the number of users.

\subsubsection{Pre-Training Performance} 

All models were pre-trained using the identical strategy on datasets with the system configurations of \((K=6, N_{\rm T}=12)\) and \((K=8, N_{\rm T}=16)\), with the results shown in Fig. \ref{fig:re:pertrain}. Models without attention mechanisms, such as MLP, CNN, and GCN, perform poorly across all tasks. In contrast, attention-based models like GAT, BERT, and UBERT show satisfactory performance, highlighting the importance of attention mechanisms in enhancing model expressiveness.

However, the performance of GPT, based on a unidirectional attention mechanism, is suboptimal. This is because it can only rely on preceding context, making it suitable for NLP token prediction tasks but not for the beamforming design tasks, where the CSI of all users is available.

The BERT incurs a performance loss of no more than $2\%$ compared to SCA across three beamforming design tasks, and even outperforms SCA by $1\%$ in the EE task. Meanwhile, the UBERT shows a performance loss of no more than $3\%$ compared to SCA. Both BERT and UBERT achieve performance comparable to  traditional optimization algorithms.

Based on the observation from Fig. \ref{fig:re:pertrain}, the subsequent experiments  focus on models with strong pre-training performance, such as GAT, BERT, and UBERT.

\begin{table*}[htbp]
\centering
\caption{Performance under different power budgets \(P_{\text{Max}}\).  }
\begin{tabular}{c|c|c|c|c|c|c|c|c|c|c}
\hline
\multirow{2}{*}{$(K, N_{\rm T})$} & \multirow{2}{*}{\textbf{Model}} & \multicolumn{3}{c|}{\textbf{SR}} & \multicolumn{3}{c|}{\textbf{MR}} & \multicolumn{3}{c}{\textbf{EE}} \\
\cline{3-11}
& & $P=1$ & $P=2$ & $P=3$ & $P=1$ & $P=2$ & $P=3$ & $P=1$ & $P=2$ & $P=3$ \\
\hline
\hline
\multirow{3}{*}{(6, 12)} & GAT & 89.81\% & 84.60\% & 81.10\% & 80.66\% & 74.20\% & 70.13\% & 91.42\% & 78.32\% & 68.82\% \\
& BERT & 99.50\% & 99.27\% & 99.13\% & 98.47\% & 98.32\% & 98.12\% & 101.24\% & 91.50\% & 83.66\% \\
& UBERT & 99.63\% & 99.54\% & 99.22\% & 98.29\% & 98.55\% & 98.03\% & 97.85\% & 95.12\% & 92.08\% \\
\hline
 \hline
\multirow{3}{*}{(8, 16)} & GAT & 84.57\% & 77.40\% & 73.13\% & 72.29\% & 64.14\% & 59.39\% & 86.50\% & 76.08\% & 64.00\% \\
& BERT & 99.41\% & 99.27\% & 99.12\% & 98.36\% & 98.25\% & 97.99\% & 101.15\% & 94.94\% & 83.62\% \\
& UBERT & 99.60\% & 99.12\% & 99.31\% & 98.04\% & 97.66\% & 98.42\% & 97.39\% & 94.28\% & 91.35\% \\
\hline
\end{tabular}
 \begin{tablenotes}
        \footnotesize
       \item {\textbf{Note}: $P$ denotes the power budget budgets $P_{\text{Max}}$. }
\end{tablenotes}
\label{pertrain:table:difPow}
\end{table*}

\subsubsection{Adaptability to Different Power Budgets}

The performance of GAT, BERT, and UBERT under different power budgets \(P_{\text{Max}} \in \{1, 2, 3\}\) W is presented in Table \ref{pertrain:table:difPow}. For the SR and MR tasks, increasing the transmit power exerts a negligible impact on performance. Compared to SCA, BERT and UBERT incur a performance loss of no more than $1\%$ and $3\%$, respectively, much smaller than the $30\%$ loss observed in GAT.

For the EE task, higher power budgets lead to a degradation in performance. GAT’s performance drops by over $20\%$ from \(P_{\text{Max
}}=1\) W to \(P_{\text{Max}}=3\) W, while BERT and UBERT experience a performance reduction of $17\%$ and $6\%$, respectively. BERT performs well at lower power (\(P_{\text{Max}}=1\)) but exhibits limited adaptability to power variations. In contrast, UBERT demonstrates superior overall adaptability to dynamic power budgets compared with BERT.

\subsubsection{Generalization Performance}
The generalization performance evaluation primarily tests the models' performance under different power budgets and user numbers.

{\it{Generalizability to power budgets.}}
To test the models' generalization performance with respect to transmit power, we pre-trained the models with \(P_{\text{Max}} = 1\) W and evaluated their performance on unseen \(P_{\text{Max}} \in \{2, 3\}\) W. The results are presented in Table \ref{pertrain:table:gen:power}. It can be observed that in terms of overall generalization performance, BERT and UBERT outperform GAT. Specifically, for the SR and MR tasks, both BERT and UBERT achieve a generalization performance over $95\%$, while GAT only maintains  a performance of above $65\%$. In the EE task, all models exhibit worse at \(P=3\) with \((K=8, N_{\rm T}=16)\). GAT achieves $63.13\%$, while BERT and UBERT reaches $82.71\%$ and $91.02\%$, respectively.

It is worth noting that as seen in Table \ref{pertrain:table:difPow}, models pre-trained with \(P_{\text{Max}} = 1\) generalize to \(P_{\text{Max}} \in \{2, 3\}\) with minimal performance loss. For instance, at \((K=8, N_{\rm T}=16)\) with \(P_{\text{Max}}=3\), BERT’s retrained performance on the SR, MR, and EE tasks is $99.27\%$, $97.99\%$, and $83.62\%$, respectively. The corresponding generalization performance is $98.64\%$, $94.35\%$, and $82.71\%$, with performance drops of only $0.63\%$, $3.64\%$, and $0.91\%$. This strong power generalization capability may be attributed to the activation functions (\ref{act}) used in the model.

\begin{table}[ht]
\centering
\caption{Generalization performance under different power budgets $P_{\rm Max}$.}
\begin{tabular}{c|c|c|c|c|c}
\hline
$(K, N_{\rm T})$&${P}$&\textbf{Model} & \textbf{SR} & \textbf{MR} & \textbf{EE} \\
 \hline
 \hline
\multirow{6}*{(6,12)}&\multirow{3}*{2}& GAT &85.10\%&74.31\%&78.34\% \\
\cline{3-6}
~&~& BERT &98.83\%& 97.26\%&91.16\% \\
\cline{3-6}
~& ~& UBERT &99.01\%&97.02\%&94.78\% \\
\cline{2-6}
~&\multirow{3}*{3}& GAT &81.79\%&70.28\%&68.88\% \\
\cline{3-6}
 ~&~& BERT &98.07\%&96.00\%&82.93\% \\
\cline{3-6}
~&~& UBERT &98.31\%&95.71\%&91.89\% \\
 \hline
  \hline
\multirow{6}*{(8,16)}&\multirow{3}*{2}& GAT &82.03\%& 70.11\%&75.37\% \\
\cline{3-6}
~&~& BERT &98.91\%&97.95\%&94.46\% \\
\cline{3-6}
~& ~& UBERT &98.93\%&97.14\%&94.18\% \\
\cline{2-6}
~&\multirow{3}*{3}&GAT&78.38\%& 65.96\%&63.13\% \\
\cline{3-6}
 ~&~& BERT &98.15\%& 96.89\%&82.71\%\\
\cline{3-6}
~&~& UBERT &98.17\%&95.57\%&91.02\% \\
 \hline
\end{tabular}
 \begin{tablenotes}
        \footnotesize
       \item {\textbf{Note}: $P$ denotes the power budget $P_{\text{Max}}$. }
\end{tablenotes}
\label{pertrain:table:gen:power}
\end{table}

{\it Generalizability to user numbers.}
To verify the suitability of the models for scenarios with varying user numbers, we evaluate the generalization performance of models pre-trained on \((K=6, N_{\rm T}=12)\) and \((K=8, N_{\rm T}=16)\) across unseen user numbers \(K \in \{5, 7, 9\}\). The results are shown in Table \ref{pertrain:table:gen:user}. It can be observed that BERT achieves the best generalization performance, which maintains over $92\%$ performance on both SR and MR tasks. For the EE task, BERT’s generalization performance even slightly surpasses that of SCA. This is because the optimal results derived by the SCA method are constrained by the predefined convergence accuracy. In contrast, UBERT exhibits relatively weaker generalization performance in the EE task, with an average performance of approximately $91\%$.

\begin{table}[ht]
\centering
\caption{Generalization performance under different numbers of users.}
\begin{tabular}{c|c|c|c|c|c}
\hline
$(K, N_{\rm T})$&${K^{*}}$&\textbf{Model} & \textbf{SR} & \textbf{MR} & \textbf{EE} \\
 \hline
 \hline
\multirow{6}*{(6,12)}&\multirow{3}*{5}& GAT &85.20\%&77.08\%&89.06\% \\
\cline{3-6}
~&~& BERT &99.31\%&98.29\%&100.22\% \\
\cline{3-6}
~& ~& UBERT &97.72\%&96.70\%&91.20\% \\
\cline{2-6}
~&\multirow{3}*{7}& GAT &85.91\%&73.54\%&88.26\% \\
\cline{3-6}
 ~&~& BERT &97.94\%&92.45\%&100.11\% \\
\cline{3-6}
~&~& UBERT &97.79\%&93.09\%&91.07\% \\
 \hline
  \hline
\multirow{6}*{(8,16)}&\multirow{3}*{7}& GAT &82.24\%&70.58\%&84.66\% \\
\cline{3-6}
~&~& BERT &99.35\%&98.50\%&100.56\% \\
\cline{3-6}
~& ~& UBERT &98.63\%&97.08\%&91.94\% \\
\cline{2-6}
~&\multirow{3}*{9}&GAT&83.06\%&69.24\%&84.38\% \\
\cline{3-6}
 ~&~& BERT &98.42\%&94.35\%&100.42\%\\
\cline{3-6}
~&~& UBERT &98.64\%&94.35\%&91.97\% \\
 \hline
\end{tabular}
\begin{tablenotes}
        \footnotesize
       \item {\textbf{Note}: $K^{*}$ denotes the number of users in the test set. }
\end{tablenotes}
\label{pertrain:table:gen:user}
\end{table}

\subsubsection{Robustness Analysis}
\begin{figure}[ht]
\begin{center}
{\includegraphics[ width=.45\textwidth]{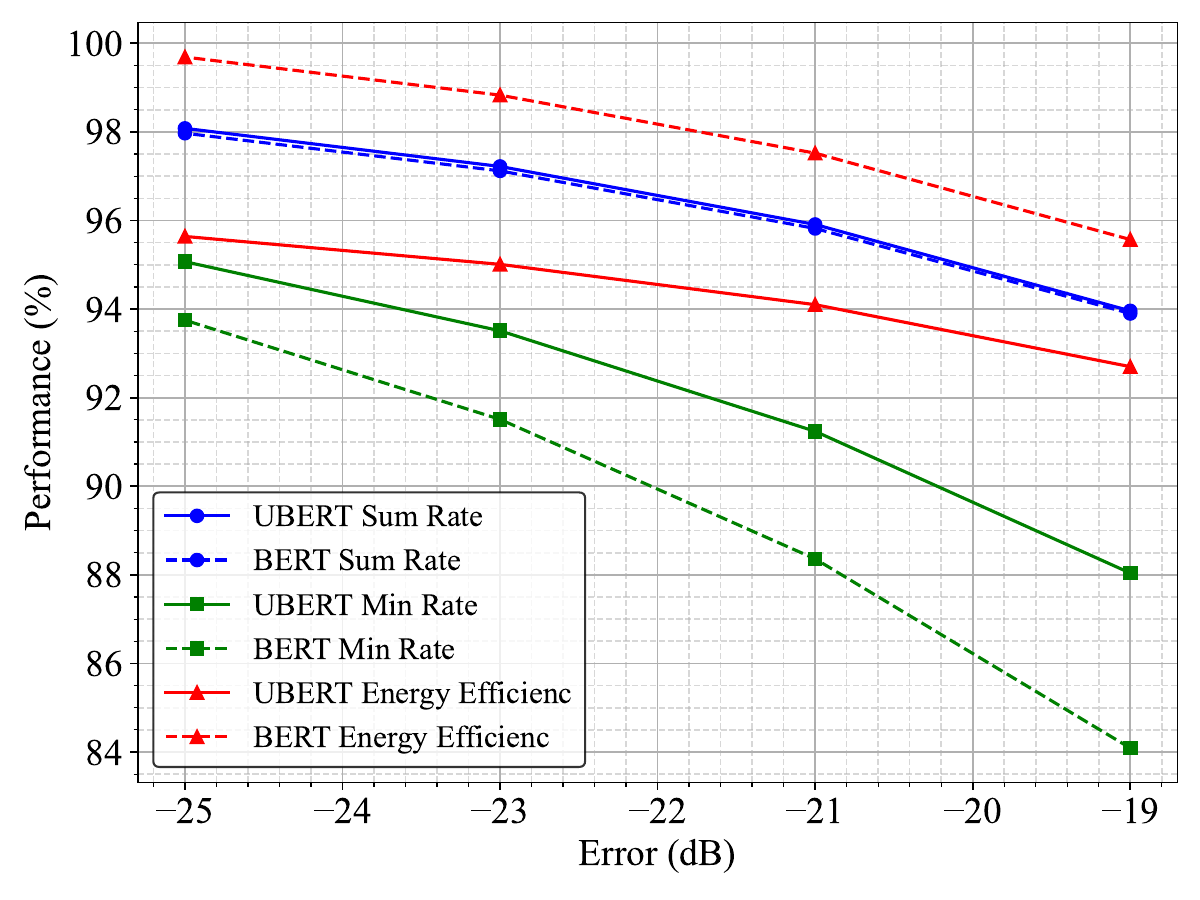}}
\caption{Performance under different CSI error levels.}
\label{pertrain:fig:csiError}
\end{center}
\end{figure}

As known, errors in the CSI can degrade transmission performance. To evaluate the robustness of the proposed approaches, CSI errors are introduced into the test samples. Specifically, for each user's CSI $\mathbf{h}_{k}$, we randomly generate a CSI error $\mathbf{e}_{k}$, where the CSI errors are assumed to be bounded and satisfy $\|\mathbf{e}_{k}\|^2=\mu\|\mathbf{h}_{k}\|^2$. Here, $\mu\in\{-25, -23, -21, -19\}$ dB denotes the CSI error level. Fig. \ref{pertrain:fig:csiError} depicts the performance of BERT and UBERT under varying CSI error levels in the system configuration of \((K=6, N_{\rm T}=12)\). 

As the CSI error level increases, the performance of both UBERT and BERT degrades gradually. For the SR task, both models exhibit strong robustness, with negligible performance differences. For the EE task, BERT demonstrates better robustness compared to UBERT. However, both models show the weakest robustness on the MR task, suffering from significant performance degradation.

\subsection{Effectiveness of Fine-Tuning Models}
In this subsection, we evaluate the fine-tuning performance of the pre-trained model on various downstream sub-tasks, which are categorized into three paradigms: scenario-adaptation fine-tuning, cross-utility fine-tuning, and few-shot fine-tuning.

\subsubsection{Scenario-Adaptation Fine-Tuning}
In the cross-scenario fine-tuning experiments, we consider two fine-tuning scenarios: one where the model is pre-trained on a small-scale system and then fine-tuned on a large-scale system, and the other where the model is pre-trained on a large-scale system and then fine-tuned on a small-scale system.

Specifically, we pre-train the models on \( (K=6, N_{\rm T}=12) \) and fine-tune them on \( (K=8, N_{\rm T}=15) \) and \( (K=8, N_{\rm T}=17) \), respectively. Additionally, we pre-train the models on \( (K=8, N_{\rm T}=16) \) and fine-tune them on \( (K=6, N_{\rm T}=11) \) and \( (K=6, N_{\rm T}=13) \), respectively. The test performance of the fine-tuned models is presented in Table \ref{fine:table:intask}. 

It is observed that when the downstream sub-tasks involve larger-scale systems, such as \( (K=8, N_{\rm T}=15) \) and \( (K=8, N_{\rm T}=17) \), both BERT and UBERT achieve a performance above $93\%$ on the SR and MR tasks. However, their performance degrades on the EE task. Specifically, BERT achieves a performance of $90.83\%$ and $81.48\%$ on the EE task for \( (K=8, N_{\rm T}=15) \) and \( (K=8, N_{\rm T}=17) \), respectively, while UBERT achieves $86.62\%$ and $88.56\%$, respectively. This performance gap may be attributed to the fact that the EE task demands more fine-tuning samples than the SR and MR tasks.

When the scale of the downstream task is smaller than that of the pre-trained model, the fine-tuned model tends to deliver superior performance. Specifically, both BERT and UBERT achieve a performance over $96\%$ on the  configurations of \( (K=6, N_{\rm T}=11) \) and \( (K=6, N_{\rm T}=13) \). This is because the pre-trained models have already captured key features from relatively large-scale systems. Consequently, when the downstream task is equal to or smaller than that of the pre-training setup, the models can achieve high performance with only a limited number of fine-tuning samples.

\subsubsection{Cross-Task Fine-Tuning}
To verify whether the MHA mechanism can learn general knowledge across different tasks, we conducted a cross-task fine-tuning experiment using the BERT model. Specifically, during the fine-tuning of the downstream task, all parameters of the transformer encoders are frozen, and only the embedding module and output layer are fine-tuned. Additionally, the sub-tasks during fine-tuning differ from those of the pre-training phase. For example, BERT is pre-trained on the SR task for \( (K=8, N_{\rm T}=16) \), and then fine-tuned on the MR and EE tasks for \( (K=6, N_{\rm T}=12) \). The experimental results are shown in Table \ref{fine:table::cross}.

\begin{table}[h]
\centering
\caption{Cross-task fine-tuning performance.}
\begin{tabular}{c|c|c|c}
\hline
{\textbf{Pre-Training }} & {\textbf{Fine-Tuning}}  & \multicolumn{2}{c}{\textbf{Fine-Tuning System}} \\
\cline{3-4}
{\textbf{ Task}}& {\textbf{Task}} &{$(6, 11)$}& {$(6, 13)$} \\
\hline
\hline
\multirow{2}{*}{SR}  & MR &95.12\%&95.96\%  \\
\cline{2-4}
& EE & 101.12\%&100.28\%\\
\hline
\hline
\multirow{2}{*}{MR}  & SR&98.79\% &98.79\% \\
\cline{2-4}
& EE &101.13\% & 100.36\% \\
\hline
\hline
\multirow{2}{*}{EE}  & SR&98.90\%&98.88\%  \\
\cline{2-4}
& MR&95.24\%&96.11\%  \\
\hline
\end{tabular}
\label{fine:table::cross}
\end{table}

It can be observed that BERT, with the attention mechanism frozen, still delivers excellent performance during cross-task fine-tuning. Its performance  exceeds $95\%$ across all downstream tasks. This indicates that the general knowledge learned by the MHA mechanism across different tasks is transferable, which further validates the feasibility of UBERT's unified architecture design.
\begin{table*}[htbp]
\centering
\caption{Fine-tuning performance.}
\begin{tabular}{c|c|c|c|c|c}
\hline
{\textbf{Pre-Training System}} & {\textbf{Model}}  &\textbf{Fine-Tuning  System}& \textbf{SR} & \textbf{MR} & \textbf{EE} \\
\hline
\hline
\multirow{4}{*}{(6, 12)}  & \multirow{2}{*}{BERT} &(8,15) & 97.76\% & 94.17\% & 90.83\%  \\
\cline{3-6}
& &(8,17)& 96.76\% & 93.66\% & 81.48\%  \\
\cline{2-6}
  & \multirow{2}{*}{UBERT} &(8,15) & 99.12\% & 96.33\% & 86.62\%  \\
\cline{3-6}
& &(8,17)& 99.10\% & 96.85\% & 88.56\%  \\
\hline
\hline
\multirow{4}{*}{(8, 16)}  & \multirow{2}{*}{BERT} &(6,11) &98.77 \% &96.87\% &100.11\%  \\
\cline{3-6}
& &(6,13)& 98.83\% & 97.26\% & 100.14\%   \\
\cline{2-6}
  & \multirow{2}{*}{UBERT} &(6,11) & 99.39\% & 97.17\% &96.77\%  \\
\cline{3-6}
& &(6,13)& 99.50\% & 98.01\% & 97.20\%  \\
\hline
\end{tabular}
\label{fine:table:intask}
\end{table*}

\subsubsection{Few-Shot Fine-Tuning}
\begin{figure}[ht]
\begin{center}
{\includegraphics[ width=.48\textwidth]{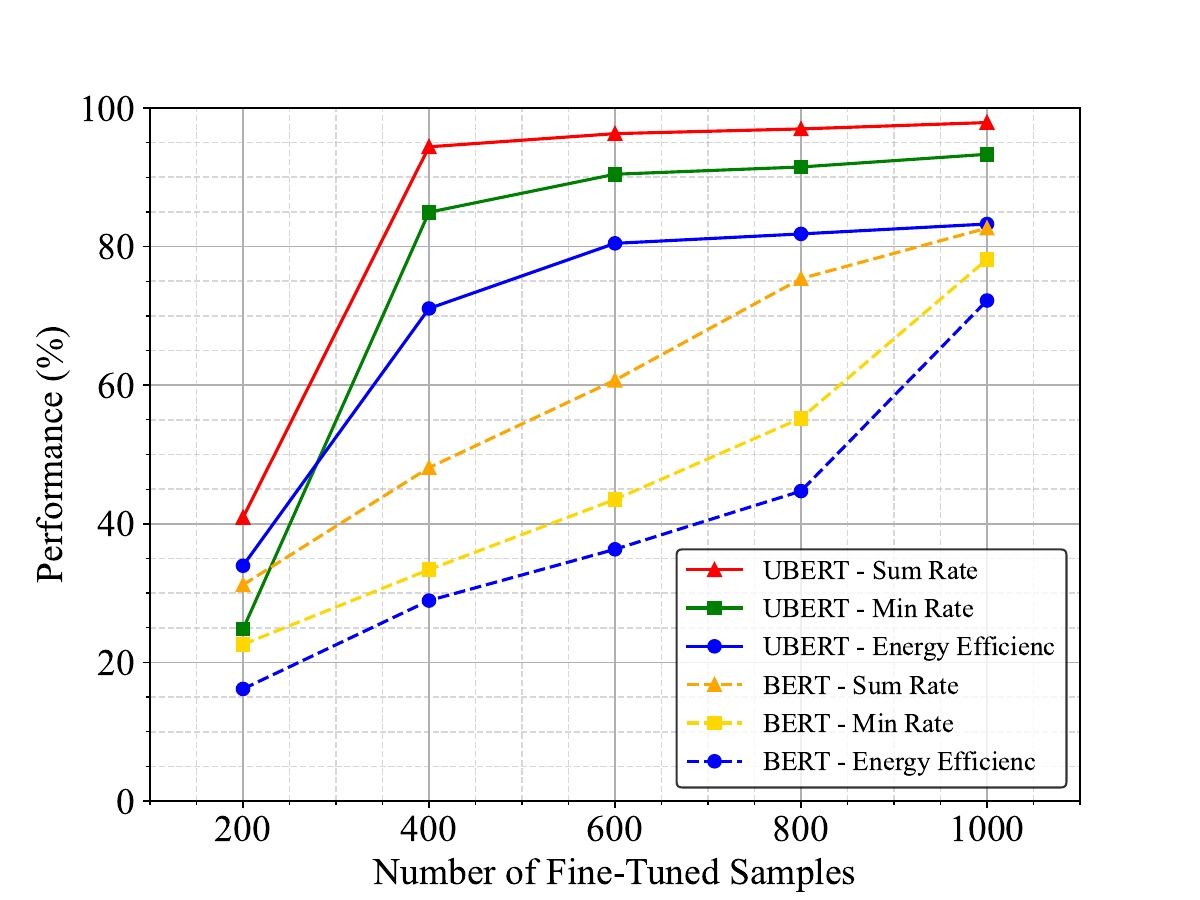}}
\caption{Fine-tuning performance under limited data scenarios.}
\label{fine:fig:few}
\end{center}
\end{figure}

To evaluate fine-tuning performance under limited data scenarios, we fine-tune the pre-trained BERT and UBERT models on \( (K=6, N_{\rm T}=11) \) using datasets of varying sizes: $200$, $400$, $600$, $800$, and $1,000$ samples. The performance curves corresponding to different fine-tuning sample sizes are illustrated in Fig. \ref{fine:fig:few}.

It can be observed that the performance of both BERT and UBERT improves steadily as the number of fine-tuning samples increases. Overall, UBERT demonstrates superior performance on small-scale datasets, which can be attributed to two key factors: first, UBERT can effectively leverage cross-task feature representations compared to BERT; second, the decoupling of input and output dimensions from system scale in UBERT helps preserve the knowledge acquired during pre-training. Notably, when the sample size reaches $600$, UBERT achieves a performance over $80\%$  across all three tasks, while BERT exceeds $65\%$ on all tasks.

\subsection{Ablation and Comparison Studies}
This subsection presents ablation and comparative experiments on the proposed loss function, model architecture, and training algorithm.

\subsubsection{Comparison of Single-Task Pre-Training Loss}
A commonly used loss function in DL methods is to directly adopt the negative of the objective function as the direction for gradient descent, refer to {\it negative objective}. Although this loss function is unsupervised and does not require a large number of labeled samples, its performance is inferior to the proposed loss function, as shown in Equation (\ref{fun:loss}). To validate our approach, we trained BERT on a small-scale dataset with \( (K = 6, N_{\rm T} = 12) \) and \( (K = 8, N_{\rm T} = 16) \), and the results are presented in Table \ref{ablation:table:singleloss}.

\begin{table}[htbp]
\centering
\caption{Performance comparison of single-task loss functions.}
\begin{tabular}{c|c|c|c}
\hline
\multirow{2}{*}{\textbf{Loss-Type}} & \multirow{2}{*}{\textbf{Task}}  & \multicolumn{2}{c}{\textbf{System}} \\
\cline{3-4}
& &{$(6, 12)$}& {$(8, 16)$} \\
\hline
\hline
\multirow{3}{*}{Proposed (\ref{fun:loss})}  & SR &94.75\%&92.51\% \\
\cline{2-4}
& MR &88.72\%&82.30\%  \\
\cline{2-4}
& EE &94.15\%&91.33\% \\
\hline
\hline
\multirow{3}{*}{{Negative Objective}}  & SR&91.87\%&89.67\% \\
\cline{2-4}
& MR &46.36\% &62.68\% \\
\cline{2-4}
& EE & 94.11\%&91.51\%\\
\hline
\end{tabular}
\label{ablation:table:singleloss}
\end{table}
It can be observed that although the negative objective function loss achieves satisfactory performance on the SR and EE tasks, a noticeable performance gap still exists between this loss function and the proposed one. Furthermore, on the MR task, the negative objective function exhibits a significant performance drop of approximately $40\%$, whereas the proposed loss function only incurs a marginal degradation performance degradation about $5\%$. This demonstrates that the proposed loss function has stronger generalizability.

\begin{table}[htbp]
\centering
\caption{Performance comparison of multi-task loss functions.}
\begin{tabular}{c|c|c|c}
\hline
\multirow{2}{*}{\textbf{Loss-Type}} & \multirow{2}{*}{\textbf{Task}}  & \multicolumn{2}{c}{\textbf{System}} \\
\cline{3-4}
& &{$(6, 12)$}& {$(8, 16)$} \\
\hline
\hline
\multirow{3}{*}{Proposed (\ref{UBert_pre})}  & SR &98.90\%&98.83\% \\
\cline{2-4}
& MR &96.55\%&95.66\%  \\
\cline{2-4}
& EE &92.93\%&91.74\% \\
\hline
\hline
\multirow{3}{*}{Sum Loss \eqref{sumloss}}  & SR&9.31\%&3.70\% \\
\cline{2-4}
& MR &1.61\% &2.26\% \\
\cline{2-4}
& EE & 8.62\%&3.68\%\\
\hline
\end{tabular}
\label{ablation:table:mutiloss}
\end{table}

\subsubsection{Comparison of Multi-Task Pre-Training Loss}
To verify the effectiveness of the proposed multi-task pre-training loss function, we conduct a comparative analysis against one of the most widely adopted multi-task loss functions, referred to as {\it sum loss} in our experiments. The sum-loss formulation is given as follows:
\begin{equation}\label{sumloss}
    \mathcal{L}_{\text{SumLoss}} = - {\rm U}_{\text{EE}}(\mathbf{W}_{\text{out}}) - {\rm U}_{\text{SR}}(\mathbf{W}_{\text{out}}) - {\rm U}_{\text{MR}}(\mathbf{W}_{\text{out}}),
\end{equation}
where \( {\rm U}_{\text{EE}}, {\rm U}_{\text{SR}}\) ,and \({\rm U}_{\text{MR}} \) denote the system utility for EE, SR, and MR, respectively. The comparison results are presented in Table \ref{ablation:table:mutiloss}.
It can be observed that the model trained with  \eqref{sumloss} fails to converge in the multi-task setting. This is primarily due to two key reasons: (1) conflicting gradient directions arising from the optimization objectives of different tasks, and (2) significant differences in the magnitude of losses across tasks, which hinder the effective optimization of shared model parameters.

\subsubsection{Ablation Experiment on UBERT Architecture}
To validate the effectiveness of the position embedding and task embedding in the proposed UBERT, we conducted a  series of ablation experiments. In these experiments, \( \text{UBERT}^{*} \) denotes the model without position embedding, and \( \text{UBERT}^{\dagger} \) represents the model without task embedding. The experimental results are presented in Table \ref{ablation:table:UBERT}. 

\begin{table}[h!]
\centering
\caption{Ablation results of UBERT components.}
\begin{tabular}{c|c|c|c}
\hline
\multirow{2}{*}{\textbf{Model}} & \multirow{2}{*}{\textbf{Task}}  & \multicolumn{2}{c}{\textbf{System}} \\
\cline{3-4}
& &{$(6, 12)$}& {$(8, 16)$} \\
\hline
\hline
 \multirow{3}{*}{UBERT} &SR&99.63\%&99.60\% \\
 \cline{2-4}
 &MR &98.29\%&98.04\%  \\
  \cline{2-4}
 &EE&97.85\%&97.39\% \\
\hline
\hline
 \multirow{3}{*}{UBERT$^{*}$} & SR&40.01\%&38.44\% \\
  \cline{2-4}
 &MR &28.15\% &27.35\% \\
  \cline{2-4}
 &EE& 35.71\%&33.68\%\\
\hline
\hline
 \multirow{3}{*}{UBERT$^{\dagger}$} & SR&79.64\%&78.89\% \\
  \cline{2-4}
 &MR &57.44\% &53.95\% \\
  \cline{2-4}
 &EE& 81.33\%&80.16\%\\
\hline
\end{tabular}
\label{ablation:table:UBERT}
\end{table}

It can be observed that the performance of UBERT degrades to varying degrees when either the position embedding or task embedding module is removed. This is because the position embedding enables UBERT to capture the relative positioning of antennas, thereby enhancing its ability to model the relative relationships between antennas. On the other hand, the task embedding enables UBERT to distinguish between different tasks in a multi-task learning setting.

\subsubsection{Comparison of Training Strategies}
\begin{figure}[ht]
\begin{center}
{\includegraphics[ width=.48\textwidth]{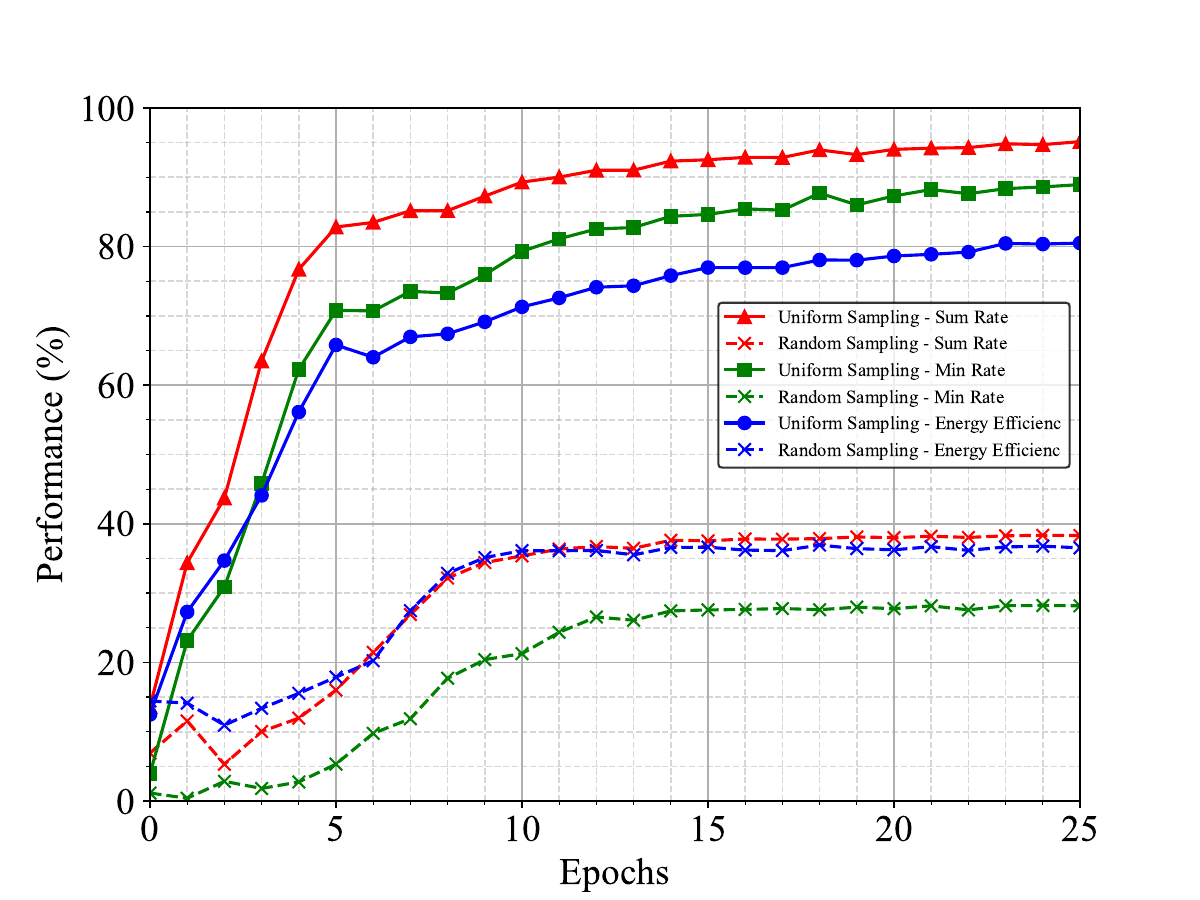}}
\caption{Effectiveness of uniform sampling in multi-task training.}
\label{ablation:fig:train}
\end{center}
\end{figure}

To validate the effectiveness of the proposed uniform sampling training algorithm, we conduct a comparative analysis against the conventional random sampling algorithm.  The variation curves of the performance metrics across different tasks during the training process are illustrated in Fig. \ref{ablation:fig:train}.

It can be observed that uniform sampling enables the model to converge more quickly and smoothly, with the final performance being significantly superior to that of random sampling. Notably, random sampling induces performance oscillations in the early training epochs, which may be due to the imbalance in gradient magnitudes across tasks, ultimately leading to poor model performance after convergence.

\section{Conclusion}

This paper has investigated the application of  pre-trained AI models in wireless communications by proposing the BERT4beam framework. We have considered beamforming tasks at the token level, enabling CSI tokenization to make it compatible with LLM architectures. Then, we have proposed two models, i.e., BERT and UBERT, to handle the token sequence-based beamforming design tasks. After pre-training, the BERT model can be directly deployed in systems with varying user counts and efficiently fine-tuned to adapt to new tasks. The UBERT model, augmented by element-wise tokenization, antenna encoding, task embedding, and pre-training with a tailored multi-task loss function, can achieve near-optimal performance across various tasks. Numerical results have demonstrated the effectiveness of the proposed approaches, and highlighted tthe superior generalization capability of LLMs as well as their promising deployment potential in practical wireless networks. {Future research will focus on extending the proposed framework to develop LLMs for various physical-layer tasks, enabling more generalizable and adaptive AI-driven strategies for wireless communication systems.}

\normalem
\bibliographystyle{IEEEtran}
\bibliography{IEEEabrv,ref}

\end{document}